\documentclass[sigconf]{acmart}

\usepackage{adjustbox}
\usepackage{booktabs}
\usepackage{multirow}
\usepackage{xcolor}
\usepackage{tikz}
\usetikzlibrary{arrows.meta, positioning, shapes.geometric}

\renewcommand\footnotetextcopyrightpermission[1]{} % removes footnote with conference information in first column

\setcopyright{none}

\acmConference[SBES 2026]{40th Brazilian Symposium on Software Engineering}{September 8--11, 2026}{São Paulo, SP, Brazil}

\AtBeginDocument{
    
}

\hypersetup{hidelinks}

\begin{document}

%% allowing the author to define a "short title" to be used on the page 
%% headers.
\title{Useful Learning Experiences: A Qualitative Study of Corporate Training in Brazilian Software Engineering}

\author{Rodrigo Siqueira}
\orcid{0009-0004-6755-9746}
\affiliation{
  \institution{CESAR School}
  \city{Recife}
  \state{PE}
  \country{Brazil}
}
\email{pinipa10@gmail.com}

\author{Antonio Oliveira}
\affiliation{
  \institution{CESAR School}
  \city{Recife}
  \state{PE}
  \country{Brazil}
}
\email{aaspo@cesar.school}

\author{Breno Alves de Andrade}
\affiliation{
  \institution{CESAR School}
  \city{Recife}
  \state{PE}
  \country{Brazil}
}
\email{baa3@cesar.school}

\author{Lidiane C S Gomes}
\affiliation{
  \institution{CESAR School}
  \city{Recife}
  \state{PE}
  \country{Brazil}
}
\email{lcsg@cesar.school}

\author{Danilo Monteiro Ribeiro}
\orcid{0000-0001-7393-729X}
\affiliation{
  \institution{CESAR School}
  \city{Recife}
  \state{PE}
  \country{Brazil}
}
\email{danilomonteiroo@gmail.com}

%% By default, the full list of authors will be used on the page
%% headers. This list is often too long and will overlap
%% other information printed in the page headers. 
%% This command allows the author to define a more concise list
\renewcommand{\shortauthors}{Siqueira et al.}
\renewcommand{\shorttitle}{Useful Learning Experiences}

%% The abstract is a short summary of the work the paper presents.
\begin{abstract}
\textbf{Context:} Quantitative studies can identify statistical predictors of training quality, but they often fail to capture what professionals themselves consider genuinely useful learning experiences and why.
\textbf{Objective:} This study qualitatively investigates which types of learning experiences are perceived as most useful by Brazilian software engineering professionals and what characteristics define this usefulness.
\textbf{Method:} Open-ended responses from 195 software engineering professionals were analyzed using Thematic Analysis \cite{terry2017thematic}, supported by frequency and lemmatization analysis using IRAMUTEQ \cite{ratinaud2009iramuteq} and co-occurrence analysis between themes.
\textbf{Results:} Five themes emerged: Continuous Technical Updating (T1), Practical and Applied Learning (T2), Formal Academic Education (T3), Social Learning and Networking (T4), and Leadership Development and Soft Skills (T5). Technical updating and practical application dominate professionals' accounts. Formal education, social learning, and soft skills are also valued as complementary dimensions.
\textbf{Conclusions:} Perceived usefulness is strongly tied to alignment with daily work demands and immediate applicability. The convergence of technical updating (T1) and practical application (T2) in both frequency and co-occurrence reinforces the imperative of continuous learning in software engineering. Useful learning is not reducible to a single modality: genuinely valued experiences span technical, academic, social, and self-directed dimensions. Formal academic education and practical learning appear to operate as complementary layers rather than competing alternatives, a reading we advance as a hypothesis for future testing. Organizations should design training ecosystems that integrate these dimensions rather than delivering isolated events.
\end{abstract}

%% the presented work. Separate the keywords with commas.
\keywords{Corporate Training, Software Engineering, Learning Experiences, Thematic Analysis, Training Usefulness, Practical Learning, Brazilian Software Industry, Qualitative Study}

%% This command processes the author, affiliation, and title
%% "frenchspacing" avoids an additional space after a period at the end of a sentence.
\frenchspacing
\maketitle

\section{Introduction}\label{section:introduction}

Corporate training in software engineering represents a growing strategic investment, yet the participant's perspective on what constitutes a genuinely useful learning experience remains underexplored \cite{de2025mapping}. The Human Resources literature has historically focused on organizational policies and practices, dedicating insufficient attention to how professionals themselves describe their most valued learning encounters \cite{santos2003employee}.

Understanding this perspective matters for several reasons. Training initiatives in the software sector may rest on generic assumptions without evidence on what to teach and how to ensure effective application in practice \cite{de2025mapping}. The effectiveness of instructional design depends on aligning content with learners' authentic needs \cite{coverstone2003training}. And the rapid pace of technological change creates a landscape in which professionals must constantly navigate between formal, informal, and self-directed learning paths \cite{diniz2024skill}.

Recent evidence reinforces this gap. \citet{de2025mapping} analyzed 26 primary studies on corporate training in software engineering, classified according to Salas' framework, and found a concentration of research on instructional methods, while critical dimensions such as Job/Task Analysis, Simulation-Based Training, and Transfer of Training remain largely unexplored. Most notably, no study focused on professionals' own narratives of what makes training useful. This gap motivates the present study.

This study addresses these gaps through a qualitative investigation of open-ended responses from Brazilian software engineering professionals, guided by the following research question:

\begin{quote}
    \textbf{RQ ---} What types of learning experiences are perceived as most useful by software engineering professionals, and what characteristics define this usefulness?
\end{quote}

The study offers five main contributions:

\begin{enumerate}
    \item Qualitative empirical evidence on the types of learning experiences valued by software engineering professionals in Brazil, grounded in their own narratives.
    \item Five thematic categories of useful learning and one emergent observation, revealing the co-existence of technical, academic, social, and self-directed dimensions.
    \item Evidence, through both thematic and co-occurrence analyses, suggesting that professionals tend to perceive formal academic education and practical learning as complementary layers rather than competing alternatives, a reading we advance as a hypothesis for future testing.
    \item Converging frequency and co-occurrence evidence that continuous technical updating and its practical application are perceived as inseparable, corroborating the imperative of lifelong learning in the field.
    \item An emergent voice of skepticism toward corporate training, pointing to tensions between organizational and individual learning agendas.
\end{enumerate}

The remainder of this article is organized as follows: Section~2 presents the theoretical framework and related work; Section~3 describes the methodology; Section~4 presents the results; Section~5 discusses the findings; Section~6 addresses limitations; and Section~7 concludes.

\section{Background}\label{section:background}

\subsection{Salas' Framework}\label{subsection:salas-framework}

We adopted the framework by \citet{salas2001science}, revised by the authors in 2012 \cite{salas2012science}, as an interpretive lens to situate professionals' narratives. The model holds that training effectiveness results from the interaction between cognitive, motivational, and organizational factors, shifting training from an isolated event to an ecosystem within the organizational context. It is a widely used framework in high-stakes domains such as aviation \cite{salas1999does} and medicine \cite{salas2012science}, where the articulation between instructional design, organizational support, and transfer to the workplace is a precondition for safe performance.

The framework is organized into four sequential components that define the company's strategy, the individual's characteristics, the instructional methods and strategies, and the post-training phase, focused on evaluation and transfer. Each of them offers interpretive hooks for the themes identified in this study.

The needs analysis answers three questions: where to train, what to teach, and whom to train. In its organizational dimension, it examines the alignment between training objectives and the company's goals, resources, and constraints. In its task dimension, it identifies the Knowledge, Skills, and Attitudes (KSAs) required for effective performance.

The antecedent conditions refer to what the participant brings and to the context preceding training. They include individual characteristics such as cognitive ability, self-efficacy, and goal orientation; the motivation to learn, understood as direction, effort, and persistence; and the pretraining environment, in which the way the organization frames the initiative shapes expectations and readiness.

The instructional methods and strategies are organized around four principles: presenting content, demonstrating the KSAs, providing practice, and delivering feedback. On this basis articulate specific learning approaches, learning technologies and distance training, simulation and games, and team training.

The post-training phase articulates two concepts. Evaluation, anchored in the typology of \citet{kirkpatrick1970evaluation} and expanded by \citet{kraiger1993application} to encompass cognitive, affective, and behavioral outcomes, assesses what was learned. Transfer refers to the application, generalization, and maintenance of KSAs in the workplace, conditioned by the organizational climate and the support of peers and supervisors.

\subsection{Related Work}\label{subsection:related_work}

\citet{coverstone2003training} argues that effective training must integrate different learning theories, combining practical exercises and collaborative activities that promote social engagement. This aligns with a view of learning as inherently situated in professional practice.

\citet{bandura1977social}'s Social Learning Theory provides a theoretical explanation for why the social dimension of learning is effective. The theory posits that individuals can acquire complex behaviors by observing others and the consequences of their actions, without relying solely on direct trial-and-error experience. This process, known as observational learning or modeling, operates through four subprocesses: attention to the model's behavior, retention of observed patterns in symbolic form, motor reproduction in one's own practice, and motivation through anticipated or indirect reinforcement (i.e., observing the consequences experienced by others). In professional contexts, these mechanisms manifest when less experienced practitioners observe how senior colleagues solve problems, internalize their reasoning patterns, and apply them in their own work.

The misalignment between formal education and software sector demands has been identified by \citet{diniz2024skill} as a leading cause of the skill gap. Yet empirical accounts of how professionals themselves perceive formal education, particularly in relation to practical learning, remain scarce.

\citet{facteau1995influence} demonstrate that organizational and social support exerts a stronger influence on training transfer than extrinsic incentives. \citet{santos2003employee} extend this by arguing that post-training behavioral change is shaped by employees' perceptions of managerial support, highlighting the role of organizational context in making learning meaningful.

Finally, \citet{salas2012science} defend learning models anchored in real work experiences and operational simulations, underscoring the centrality of practical application, a theme that recurs prominently in participants' narratives.

Taken together, the studies above approach training mostly from the organization's or the instructional designer's standpoint: what to deliver, how to support transfer, and which conditions improve effectiveness. They establish that applicability, social support, and formal foundations matter, but they tend to infer these from policies, interventions, or measured outcomes rather than from how professionals themselves narrate their most useful learning. The advance of the present study is twofold. First, we supply first-person, field-specific evidence that complements this predominantly outcome-based and quantitative literature. Second, we surface a distinction that prior work does not foreground, namely that the much-cited industry--academia skill gap \cite{diniz2024skill} need not coincide with how the individual professional experiences formal education, which in our sample is perceived as a complementary foundation rather than a competitor to practice. A reader interested in designing training ecosystems may therefore gain a practitioner-centered perspective that intervention- and policy-centered studies do not directly offer.

\section{Method}\label{section:method}

\subsection{Study Design}
This study adopts a qualitative design grounded in Thematic Analysis \cite{terry2017thematic}. Data were collected through open-ended survey questions administered to Brazilian software engineering professionals. Initial codes were generated from the data rather than from a predefined coding scheme; however, theme construction was inevitably informed by the research team's prior familiarity with Salas' framework. We therefore characterize the process as predominantly inductive in its coding stage but abductive overall, moving between the data and the interpretive lens of Section~\ref{subsection:salas-framework} rather than claiming theory-free induction. The implications of this stance are discussed further in Section~\ref{sec:limitations}. The qualitative and quantitative data were collected simultaneously through the same instrument, but the qualitative strand was analyzed and reported independently.

\subsection{Instrument Design}

The questionnaire included two open-ended questions designed to complement structured Likert-scale items, allowing participants to express perceptions not captured by closed items. This study analyzes only one of them, focused on the most valued learning experiences:

\begin{quote}
    \textbf{QQ1:} ``What was the most useful learning experience you have had? Why?''
\end{quote}

The instrument was developed iteratively, with a pilot study involving 10 professionals assessing clarity, ambiguity, relevance, and completion time. Based on feedback, the term \textit{training} was replaced with \textit{learning experience} to broaden the scope of responses.

\subsection{Participants and Sampling}

The target population comprises professionals working in software engineering roles in Brazil, including software development, quality assurance, technical leadership, software architecture, DevOps, data engineering, and people/project management within software teams. Recruitment was by convenience through LinkedIn, WhatsApp and Telegram groups, and email lists. Participation was voluntary, with no financial incentives.

Inclusion criteria:
\begin{itemize}
    \item being 18 years of age or older;
    \item working professionally in the software engineering area;
    \item having participated, in the last 12 months, in at least one corporate training sponsored by the employing organization.
\end{itemize}

Of 282 valid responses to the full questionnaire, \textbf{195} provided a valid response to QQ1. This reduction reflects the non-mandatory nature of the open-ended question. These 195 responses constitute the analytical corpus. Data collection took place entirely online, from September 9 to November 2, 2025 ($\approx$8 weeks). No data imputation was performed.

\subsection{Data Analysis}

Three complementary analytical procedures were applied.

\textbf{Frequency and lemmatization analysis} was performed using IRAMUTEQ \cite{ratinaud2009iramuteq}, an open-source software for textual data analysis. Lemmatization is the linguistic procedure of reducing inflected words to their base form (e.g., \textit{trained}, \textit{training}, \textit{trains} all become \textit{train}), allowing frequency counts to capture semantic regularities rather than surface variations.

\textbf{Thematic Analysis} followed the six-stage procedure by \citet{terry2017thematic}: (1)~familiarization; (2)~initial coding; (3)~theme generation; (4)~theme review; (5)~definition and naming; and (6)~report production. A single response could be associated with more than one theme.

\textbf{Co-occurrence analysis} examines how often pairs of categories appear together in the same unit of analysis. After theme consolidation, a co-occurrence was recorded whenever two themes were assigned to the same participant's response, and the resulting frequencies were used to construct a network graph in which nodes represent themes and edges represent shared responses.

\textbf{Saturation assessment.} To assess the stability of the resulting categories, we examined saturation by tracking the cumulative emergence of codes and themes across responses, processed in coding order. For each successive response we recorded whether it introduced any code or theme not previously seen, producing two novelty curves. This indicator does not replace inter-rater agreement, but it makes transparent and replicable the point at which the corpus stopped yielding new categories; the underlying assignments are available in the replication package. Results are reported in Section~\ref{sec:thematic}.

Table~\ref{tab:trace_example} traces a single response through the six stages, illustrating how raw data were progressively segmented into codes and consolidated into themes.

\begin{table*}[ht]
    \centering
    \caption{Example of the thematic analysis process applied to a single response (P273)}
    \label{tab:trace_example}
    \small
    \renewcommand{\arraystretch}{1.5}
    \begin{tabular}{| p{0.18\linewidth} | p{0.74\linewidth} |}
        \hline
        \textbf{Phase} & \textbf{Result} \\
        \hline
        1 -- Familiarization &
        P273: ``Leadership training. Because it connected a personal desire with the company's needs, and the content was genuinely very good.'' \\
        \hline
        2 -- Initial coding &
        Three codes assigned: \textit{Training}; \textit{Leadership}; \textit{Problem-solving} (connecting personal goals with organizational needs). \\
        \hline
        3 -- Theme construction &
        Codes associated with three candidate axes: Courses and Training (\textit{Training}); Leadership and Soft Skills (\textit{Leadership}); Practical and Applied Learning (\textit{Problem-solving}). \\
        \hline
        4 -- Review and refinement &
        Coherence verified: \textit{Training} aligns with the technical updating axis; \textit{Leadership} with the behavioral competencies axis; \textit{Problem-solving} with the practical application axis. All three confirmed as semantically distinct. \\
        \hline
        5 -- Definition and naming &
        T1 -- Continuous Technical Updating; T2 -- Practical and Applied Learning; T5 -- Leadership Development and Soft Skills. \\
        \hline
        6 -- Report production &
        Response used as illustrative quote under T5 (Section~\ref{sec:thematic}); generated co-occurrence records between T1--T2, T1--T5, and T2--T5 (Section~\ref{sec:cooccurrence}). \\
        \hline
    \end{tabular}
\end{table*}

The coding was carried out by a single researcher, under the regular supervision of the senior author, who reviewed the candidate axes and their consolidation into final themes. While single-coder analysis remains a recognized limitation (see Section~\ref{sec:limitations}), the systematic protocol was followed rigorously to enhance analytical discipline and transparency.

The researcher responsible for the coding has ten years of experience in software engineering, including two years as a technical leader, a role that involves fostering corporate education within the team. The researcher is also a graduate student in software engineering, establishing a direct academic engagement with the topic. This dual profile, professional and academic, offers familiarity with respondents' vocabulary and practices, but also introduces potential biases.

The most relevant of these is prior familiarity with Salas' framework, which is shared by the entire research team and could not be neutralized. We therefore make this influence explicit: although initial codes were derived from participants' own wording, the subsequent grouping of codes into axes was shaped by the team's exposure to the framework, so that categories compatible with Salas' vocabulary (for example, the separation between needs analysis, instructional methods, and transfer) may have been more readily perceived than alternative organizing schemes. This is why we describe the overall process as abductive rather than purely inductive. As a modest, descriptive triangulation, the IRAMUTEQ frequencies point in the same direction as the coding without being derived from it: \textit{course} and \textit{training} are among the most frequent nouns and correspond to the dominant theme T1, while \textit{practice} is the single most frequent verb in the corpus and names the second theme, T2 (Table~\ref{tab:formas_frequencia}). We do not test a statistical association between word counts and theme assignments; we note only that the vocabulary participants used most often echoes the two most salient themes, which suggests that their prominence is reflected in the raw text and not only in the coder's reading. The raw responses and all intermediate artifacts are released so that readers can inspect, contest, or re-derive the category structure.

\subsection{Translation of Participant Quotations}

The survey was administered in Brazilian Portuguese, and all participant responses were originally written in Portuguese. The quotations presented in this paper were translated into English with the assistance of Generative AI tools, as declared in the Declaration of the Use of Artificial Intelligence. In all cases, the translations were reviewed by the authors to ensure that the participant's intended meaning was preserved.

\subsection{Ethical Considerations}

The study was approved by the Research Ethics Committee under CAEE 91121125.4.0000.5208 (Opinion No.\ 7.816.810).
Participation was voluntary, anonymous, and confidential. Participants could withdraw at any time. The main risks considered were data leakage and psychological discomfort from recalling past experiences, both mitigated according to the protocol approved by the ethics committee.

\section{Results}\label{section:results}

\subsection{Participant Profile}
\label{sec:qs}

The 195 respondents are predominantly male (73.8\%), mean age 34.2 years ($SD=7.7$), highly educated (91.3\% with at least a complete undergraduate degree; 59.0\% with postgraduate qualifications). Senior is the most frequent professional level (29.2\%), followed by Manager (20.5\%) and Mid-level (17.4\%). Most work in large enterprises with 100+ employees (77.9\%), and 46.7\% report more than eight years of experience. Software Development is the predominant area (39.0\%), followed by People/Project Management (16.4\%) and Quality Assurance (15.4\%). Most (71.3\%) reported their most recent training was voluntary.

\subsection{Frequency Analysis}
\label{sec:qq}

Table~\ref{tab:formas_frequencia} presents the five most salient semantic forms identified through IRAMUTEQ lemmatization.

\begin{table}[ht]
    \centering
    \caption{Most frequent semantic forms in the QQ1 corpus}
    \label{tab:formas_frequencia}
    \begin{tabular}{|l|c|c|}
        \hline
        \textbf{Word} & \textbf{Frequency} & \textbf{Grammatical class} \\
        \hline
        course        & 47 & Noun \\
        company       & 29 & Noun \\
        area          & 23 & Noun \\
        training      & 23 & Noun \\
        practice      & 21 & Verb \\
        \hline
    \end{tabular}
\end{table}

Nouns associated with professional training and organizational context, \textit{course} ($n=47$), \textit{company} ($n=29$), \textit{area} ($n=23$), and \textit{training} ($n=23$), form the semantic core of participants' discourse. The first verb to appear with high frequency is \textit{practice} ($n=21$), suggesting that practical application is perceived as central to learning usefulness.

\subsection{Thematic Analysis}
\label{sec:thematic}

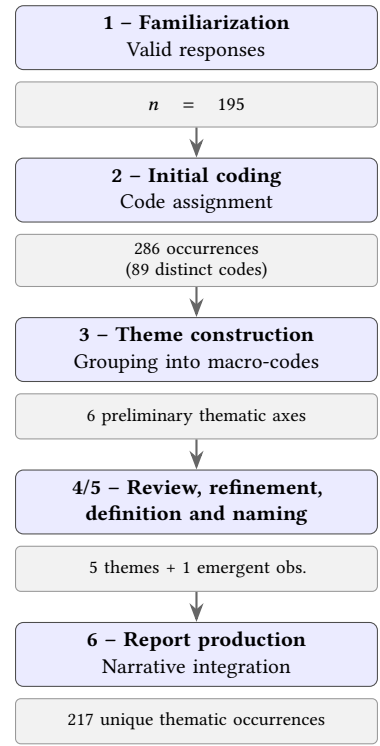
\begin{figure}[ht]
    \centering
    \begin{tikzpicture}[
        node distance=0.55cm,
        phase/.style={
            rectangle, rounded corners=3pt, draw=black!70, fill=blue!8,
            minimum width=4.8cm, minimum height=0.85cm,
            font=\small, align=center, text width=4.6cm
        },
        quant/.style={
            rectangle, rounded corners=2pt, draw=black!40, fill=gray!10,
            minimum width=4.8cm, minimum height=0.6cm,
            font=\footnotesize, align=center, text width=4.6cm
        },
        arr/.style={-{Stealth[length=2.5mm]}, thick, black!60}
    ]
        \node[phase] (p1) {\textbf{1 -- Familiarization}\\Valid responses};
        \node[quant, below=0.15cm of p1] (q1) {$n = 195$};
        \node[phase, below=0.4cm of q1] (p2) {\textbf{2 -- Initial coding}\\Code assignment};
        \node[quant, below=0.15cm of p2] (q2) {286 occurrences\\(89 distinct codes)};
        \node[phase, below=0.4cm of q2] (p3) {\textbf{3 -- Theme construction}\\Grouping into macro-codes};
        \node[quant, below=0.15cm of p3] (q3) {6 preliminary thematic axes};
        \node[phase, below=0.4cm of q3] (p4) {\textbf{4/5 -- Review, refinement,}\\\textbf{definition and naming}};
        \node[quant, below=0.15cm of p4] (q4) {5 themes + 1 emergent obs.};
        \node[phase, below=0.4cm of q4] (p6) {\textbf{6 -- Report production}\\Narrative integration};
        \node[quant, below=0.15cm of p6] (q6) {217 unique thematic occurrences};
        \draw[arr] (q1.south) -- (p2.north);
        \draw[arr] (q2.south) -- (p3.north);
        \draw[arr] (q3.south) -- (p4.north);
        \draw[arr] (q4.south) -- (p6.north);
    \end{tikzpicture}
    \caption{Phases of thematic analysis and quantitative refinement at each stage.}
    \label{fig:phases}
\end{figure}

Figure~\ref{fig:phases} summarizes the six-stage progression with quantitative outputs. From 195 responses, 286 code occurrences were generated across 89 distinct codes, grouped into 6 preliminary axes, and consolidated into \textbf{5 themes} and \textbf{one emergent observation} encompassing 217 unique thematic occurrences. The total exceeds the number of participants because a single response could map to multiple themes. The emergent observation ($n=5$) was distinguished from themes due to its low recurrence. Table~\ref{tab:frequencia_temas} presents the final themes and their frequencies.

The saturation assessment described in Section~\ref{section:method} indicates that thematic saturation was reached early: all six thematic axes (the five themes plus the emergent observation) had already emerged within the first 16 coded responses, that is, in roughly the first 8\% of the corpus. None of the remaining 179 responses (about 92\% of the corpus) introduced a new theme; they contributed only to the redistribution of frequencies and to a long tail of low-frequency, idiosyncratic codes. Code-level novelty declined steadily rather than abruptly: the majority of distinct codes had appeared by the midpoint of the corpus, after which newly introduced codes became progressively rarer. We read this as evidence of stability at the thematic level, while acknowledging that code-level saturation in the strict sense was approached but not fully exhausted, a pattern consistent with open-ended responses that occasionally surface singular experiences.

\begin{table}[ht]
    \centering
    \caption{Final themes, emergent observation, and their frequencies}
    \label{tab:frequencia_temas}
    \begin{tabular}{|l|l|c|}
        \hline
        \textbf{Code} & \textbf{Theme} & \textbf{Freq.} \\
        \hline
        T1 & Continuous technical updating & 91 \\
        T2 & Practical and applied learning & 39 \\
        T3 & Formal academic education & 35  \\
        T4 & Social learning and networking & 30 \\
        T5 & Leadership development and soft skills & 17 \\
        \hline
        EO1 & Self-directed learning and skepticism & 5 \\
        \hline
    \end{tabular}
\end{table}

\subsubsection{T1 --- Continuous Technical Updating}

Courses, certifications, and technology-specific studies dominate the corpus. Participants associate usefulness with alignment to everyday work demands and current industry trends.

\begin{quote} \textit{P9: SQL training, as it allowed me to deepen my knowledge of a technology I use daily.}
\end{quote}

\begin{quote} \textit{P12: Java certification, which changed my understanding of the language and raised the level of my solutions.}
\end{quote}

\begin{quote} \textit{P180: Introduction to AI. It helped me take the first step toward updating myself on basic concepts.}
\end{quote}

\subsubsection{T2 --- Practical and Applied Learning}

Hands-on activities, experimentation, and real-project engagement emerge as key drivers of usefulness. Participants stress that immediate application consolidates knowledge and supports development.

\begin{quote} \textit{P43: Hands-on, because you truly have the opportunity to put what you learned into practice, even if only at that moment, and this helps consolidate knowledge.}
\end{quote}

\begin{quote} \textit{P122: Learning about the market and the business rules related to the developed product.}
\end{quote}

\subsubsection{T3 --- Formal Academic Education}

Undergraduate, postgraduate, master's, doctoral, and MBA programs are associated with building solid conceptual foundations, market recognition, and expanded professional opportunities.

\begin{quote} \textit{P111: University education, because the fundamentals are essential and, without it, I would not have pursued further learning.}
\end{quote}

\begin{quote} \textit{P37: Master's degree, because it broadened my perspective on several topics and connected me with highly qualified people.}
\end{quote}

\subsubsection{T4 --- Social Learning and Networking}

Events, conferences, communities, mentoring, and peer exchanges highlight the value of collective learning, diverse perspectives, and professional network expansion.

\begin{quote} \textit{P4: Participation in events, as it allows contact with professionals from other companies and the exchange of experiences.}
\end{quote}

\begin{quote} \textit{P155: A software architecture mentoring program. I learned a lot from discussion circles and accounts from more experienced professionals.}
\end{quote}

\subsubsection{T5 --- Leadership Development and Soft Skills}

Leadership, communication, conflict management, and emotional intelligence are valued as competencies that enhance performance beyond technical mastery.

\begin{quote} \textit{P80: Emotional intelligence to deal with internal and external conflicts.}
\end{quote}

\begin{quote} \textit{P273: Leadership training, because it connected a personal desire with the company's needs, and the content was genuinely very good.}
\end{quote}

\subsubsection{EO1 --- Self-Directed Learning and Skepticism (Emergent Observation)}

A small number of responses ($n=5$) described self-initiated study and personal development planning as their most useful learning. Though too infrequent to constitute a consolidated theme, these responses carry a distinctive tone of skepticism toward corporate training, suggesting a subset of professionals who have consciously decoupled their learning agendas from organizational offerings.

\begin{quote} \textit{P5: None came from companies. I do not believe that any content produced by companies is truly of high quality or essential.}
\end{quote}

\begin{quote} \textit{P260: Creating my own strategic plan, focusing not on the company but on myself---the well-known Individual Development Plan without manipulation to serve specific interests.}
\end{quote}

Given its low frequency, this observation is reported for transparency and as a signal for future investigation.

\subsection{Co-Occurrence Analysis}
\label{sec:cooccurrence}

\begin{figure}[ht]
    \centering
    \includegraphics[width=\columnwidth]{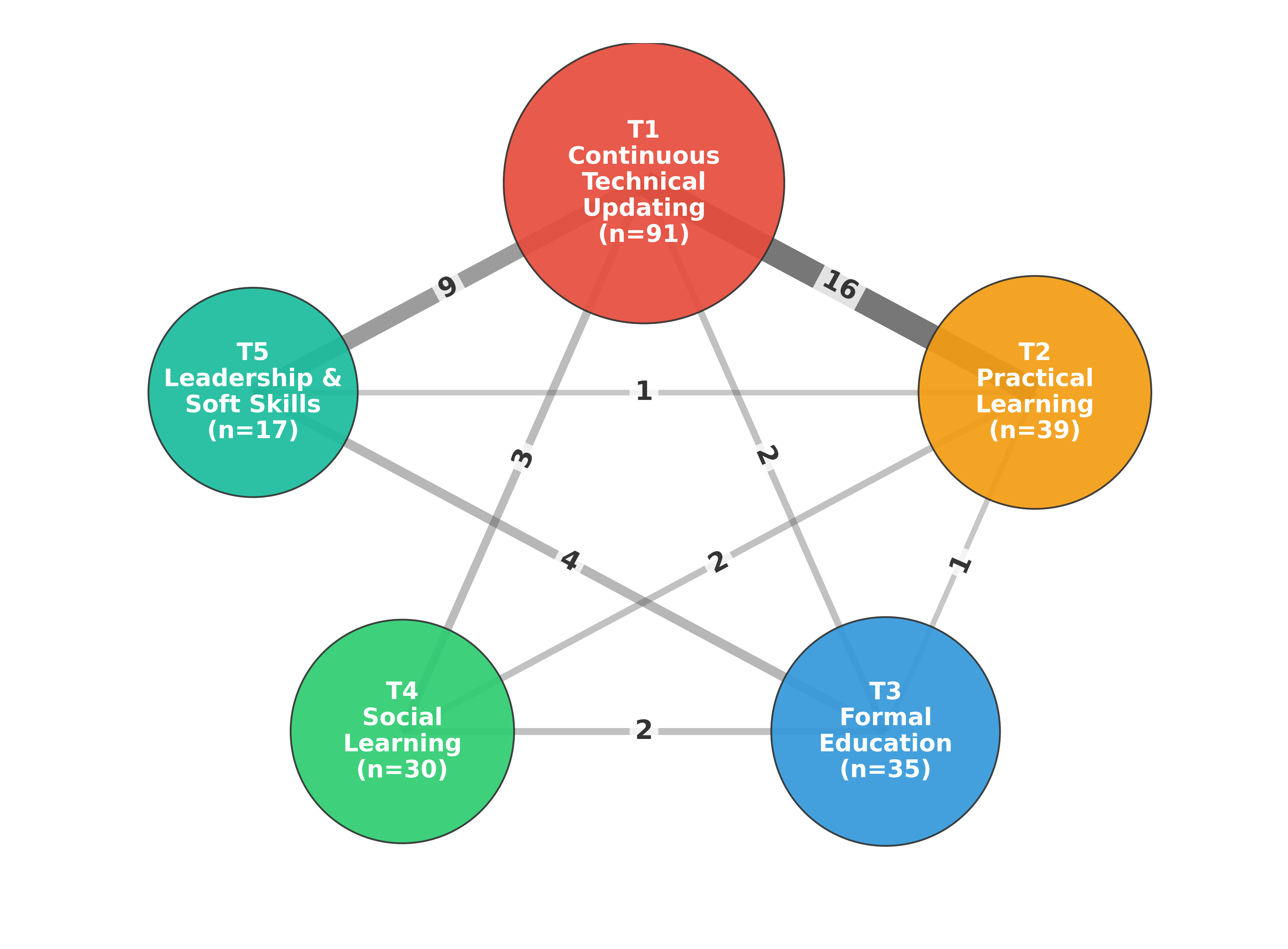}
    \caption{Co-occurrence graph between identified themes.}
    \label{fig:cooccurrence}
    \vspace{2pt}
    \raggedright\scriptsize
    \begin{tabbing}
    \hspace{0.5em}\= \\
    \> (a) Each node represents a theme, with its frequency in parentheses.\\
    \> (b) Each edge connects themes cited in the same response.\\
    \> (c) The number on the edge indicates the number of co-occurrences.\\
    \> (d) Edge thickness is proportional to the number of co-occurrences.
    \end{tabbing}
\end{figure}

Since a single response could map to multiple themes, co-occurrence analysis examined how frequently themes were cited by the same participant. Figure~\ref{fig:cooccurrence} represents the resulting network.

The strongest co-occurrence links T1 and T2, with 16 shared responses. Professionals who value technical updating frequently associate it with the need for immediate practical application, reinforcing that staying continuously updated is driven by the demand to solve everyday problems in a rapidly changing work environment. The second strongest link connects T1 and T5 (9 co-occurrences), suggesting that a subset of professionals values technical learning alongside behavioral competencies.

T3 shows moderate connections with T5 ($n=4$), T1 ($n=2$), and T4 ($n=2$), but a notably low co-occurrence with T2 ($n=1$). We interpret this with caution. A low co-occurrence is, on its own, ambiguous: it is consistent both with a \textit{complementary} reading, in which formal education and hands-on practice operate as distinct layers that participants rarely describe in the same breath, and with a more \textit{competitive} reading, in which the two are mentally filed as alternatives. Several responses point toward complementarity, framing formal education as the foundation on top of which practice is later built, as in \textit{``University education, because that is where I learned the basics of professional life.''} The single response that did bridge T2 and T3 is illuminating precisely because it is exceptional: \textit{``MBA in project management, which was completely practical and taught by professionals active in the field,''} suggesting that the pairing occurs mainly when the formal program is itself experienced as applied. The current data cannot adjudicate between the two readings, so we treat the complementary interpretation as the better-supported hypothesis and flag the relationship between formal and practical learning as a target for future investigation rather than a settled result.

The low co-occurrence between T2 and T5 ($n=1$) admits a similar reading. Soft-skills development was far more often cited alongside technical updating (T1--T5 $= 9$) than alongside hands-on practice, and the one response linking T2 and T5 was a leadership course described as both personally motivating and concretely applicable (P273). This pattern is more easily seen on the graph once the absence of an edge is read as informative, not merely as missing data.

T4 shows modest but distributed connections across T1 ($n=3$), T2 ($n=2$), and T3 ($n=2$), consistent with a cross-cutting role that enhances other forms of learning.

\section{Discussion}\label{section:discussion}

This section discusses the findings in light of the training literature, addressing \textbf{RQ:} \textit{What types of learning experiences are perceived as most useful by software engineering professionals, and what characteristics define this usefulness?}

In direct answer to the RQ, software engineering professionals perceive as most useful the learning experiences that align with daily work demands and enable immediate application. Five dimensions characterize this usefulness: continuous technical updating (T1), practical and applied learning (T2), formal academic education as a conceptual foundation (T3), social learning and networking (T4), and behavioral and leadership competencies (T5). Rather than appearing as mutually exclusive alternatives, these dimensions tend to be described as articulated layers, with the T1--T2 relationship emerging as the core of this perception of usefulness. We elaborate on the strength of the evidence for each dimension below, and treat the more interpretive readings as hypotheses rather than settled results.

T1 ($n=91$) and T2 ($n=39$) together reinforce a well-established principle: perceived usefulness is strongly tied to alignment with daily work demands and the possibility of immediate application \cite{coverstone2003training}. The co-occurrence analysis strengthens this finding: T1 and T2 share the strongest network link (16 co-occurrences), indicating that professionals perceive continuous updating and practical application as inseparable, driven by the need to remain current in a field where daily problem-solving demands immediate use of newly acquired knowledge. The high frequency of the verb \textit{practice} in the lemmatization analysis is consistent with this pattern.

T3 ($n=35$) emerged with notable prominence despite the well-documented misalignment between academia and industry \cite{diniz2024skill}. Participants consistently framed academic education as a conceptual scaffolding that made subsequent learning possible. Within Salas' framework, this aligns with the \textit{Training Needs Analysis} dimension, specifically the identification of foundational KSAs \cite{salas2001science}. This finding suggests that the skill gap debate involves three distinct actors. The tension documented by \citet{diniz2024skill} operates primarily between \textit{industry}, which demands immediate problem-solving readiness, and \textit{academia}, whose curricula struggle to keep pace. However, for the \textit{individual professional} in this sample, formal education is perceived as complementary to practical and technical learning, a foundational layer that enables absorption of successive waves of technological change throughout a career. The skill gap, then, is a tension between industry and academia that does not necessarily reflect the individual professional's experience of learning usefulness. The co-occurrence network is consistent with this reading: T3 connects moderately with T5 and T1 but co-occurs only once with T2. As discussed in Section~\ref{sec:cooccurrence}, a low co-occurrence is on its own ambiguous, so we advance the complementary interpretation as the better-supported hypothesis rather than a confirmed result.

T4 ($n=30$) highlights communities of practice, conferences, and mentoring as important learning contexts. This aligns with \citet{facteau1995influence}'s finding on the role of peer and supervisor support in training transfer, and with \citet{devaraj2004measure}'s observation on interaction opportunities. The participants' narratives also resonate with \citet{bandura1977social}'s observational learning framework: professionals who described mentoring programs (P155) and discussion circles with experienced colleagues are engaging in the subprocesses of modeling, attending to how senior practitioners approach problems, retaining those patterns, reproducing them in their own projects, and being reinforced by peer feedback and successful outcomes. In software engineering, T4 maps onto the strong culture of knowledge sharing through meetups, open-source communities, and tech talks, informal systems that operate alongside corporate training and function as natural environments for observational learning.

T5 ($n=17$) reflects an emerging recognition that behavioral competencies, communication, conflict management, and emotional intelligence, complement technical expertise, particularly for professionals in or approaching leadership roles.

EO1 ($n=5$) does not carry the same weight as the consolidated themes, but its distinctive tone of skepticism signals a tension between organizational and individual learning agendas. It also offers an interesting counterpoint to T4: while the majority value socialization and collective exchange, a small subset favors entirely self-directed paths detached from organizational offerings.

These themes acquire specific contours in the software engineering context. T1 reflects a distinctive characteristic of the field: rapid technological change imposes a continuous cycle of learning and obsolescence more acute than in most other domains \cite{diniz2024skill}. T2 resonates with practices like pair programming, code review, and on-the-job mentoring, activities that blur the boundary between training and productive work. T3's role as conceptual scaffolding is particularly relevant where foundational knowledge in algorithms, data structures, and design principles enables professionals to navigate rapid shifts rather than being rendered obsolete by each new tool.

Taken together, useful learning is not reducible to a single modality. The co-occurrence network (Figure~\ref{fig:cooccurrence}) visually reinforces this: the five themes form an interconnected structure linked through shared narratives, pointing to a need for training ecosystems rather than isolated events.

Based on these patterns, four analytical propositions are offered for future investigation:

\begin{itemize}
    \item \textbf{PR1:} Continuous technical updating and its immediate practical application are perceived as inseparable dimensions of useful learning, driven by the need to remain current in a rapidly evolving technological environment. \textit{(Supported by: T1 dominance, $n=91$; T1--T2 co-occurrence = 16; high frequency of ``practice'' in lemmatization.)}
    \item \textbf{PR2:} Formal academic education functions as a conceptual foundation that enables subsequent technical and practical learning. \textit{(Supported by: T3 prominence, $n=35$; low T3--T2 co-occurrence suggests complementary layers rather than interchangeable experiences.)}
    \item \textbf{PR3:} Social and collaborative interactions enhance the perceived value of other learning modalities. \textit{(Supported by: T4, $n=30$, shows distributed connections across T1, T2, and T3, consistent with a cross-cutting role.)}
    \item \textbf{PR4:} Perceived learning usefulness is multidimensional, integrating technical, practical, academic, and social components. \textit{(Supported by: interconnected co-occurrence structure; 217 thematic occurrences from 195 responses indicate systematic overlap.)}
\end{itemize}

These propositions do not claim confirmatory status. They are offered as empirically grounded directions for quantitative studies.

\section{Threats to Validity and Limitations}
\label{sec:limitations}

\textbf{Coding and interpretative bias}: The thematic coding was performed by a single researcher, under the supervision of the senior author, without parallel coding or formal inter-rater reliability. The research team's shared familiarity with Salas' framework means the analysis is best understood as abductive rather than purely inductive (Section~\ref{section:method}): codes were data-driven, but the construction of axes was filtered through that theoretical lens, which may have favored Salas-compatible categories over alternative framings. We did not formally measure saturation in advance; instead, we assessed it \textit{a posteriori} by tracking the emergence of new codes and themes across the corpus (Section~\ref{sec:thematic}). This assessment shows thematic stability, with no new theme appearing after the first 16 responses, while indicating that strict code-level saturation was approached but not fully reached. These are post-hoc indicators and do not replace inter-rater agreement. To mitigate the remaining risks, the six-stage protocol \cite{terry2017thematic} was followed rigorously, and the replication package is released so that the category structure can be independently inspected and re-derived. Future replications should consider multiple coders and a saturation criterion defined \textit{a priori}.

\textbf{Sampling, self-selection, and memory bias}: Recruitment via professional networks (LinkedIn, WhatsApp, Telegram) may over-represent professionals already engaged with training and professional development. The open-ended question was optional, and only 195 of 282 participants (69.1\%) responded. The 87 non-respondents may have systematically different profiles or attitudes toward training, which constitutes a non-response bias. As a partial mitigation, we observed that the sociodemographic profile of the 195 respondents closely mirrors that of the full sample of 282 in terms of gender, age, education, professional level, company size, and area of work, suggesting that the analyzed corpus is not skewed in observable demographic dimensions. However, this does not rule out non-observable differences, such as engagement with the topic or strength of opinions. Additionally, QQ1 asked about the ``most useful'' experience without distinguishing training types (onboarding, upskilling, workshops, compliance), meaning responses may mix formats with different objectives and pedagogies. Participants also described retrospective experiences, which may not correspond to the most recent, introducing memory distortion.

\textbf{Sample profile}: The sample is predominantly male (73.8\%), highly educated (91.3\% with at least a complete undergraduate degree, 59.0\% with postgraduate qualifications), and restricted to Brazil. These characteristics may not reflect the full diversity of the Brazilian software workforce and limit generalizability. The high educational profile may particularly bias the prominence of T3. Two observations temper, without eliminating, this concern. First, higher education was not an inclusion criterion, so the educational profile is an observed characteristic of those who chose to respond rather than a condition we imposed. Second, even within this highly educated group, T3 is not dominant: it ranks behind both T1 ($n=91$) and T2 ($n=39$), which is difficult to reconcile with an account in which the educational profile alone drives its salience. We therefore read T3 as a foundational layer as experienced within this sample, not as a claim about the population of Brazilian software professionals. Professional level categories (e.g., Junior, Senior, Manager) may also be interpreted heterogeneously across organizations; future studies should adopt explicit operational definitions for each level.

\textbf{Focus on software engineering}: Results are not automatically transferable to other IT specialties or sectors.

\section{Conclusion}\label{section:conclusion}

This study investigated which learning experiences Brazilian software engineering professionals perceive as most useful and what defines this usefulness. Through thematic analysis of 195 open-ended responses, supported by frequency, lemmatization, and co-occurrence analyses, five themes and one emergent observation were identified.

The convergence of Continuous Technical Updating (T1, $n=91$) and Practical and Applied Learning (T2, $n=39$), both in frequency and in co-occurrence (16 shared responses), points to a central finding: in a field where technologies evolve on timescales of months, professionals perceive continuous updating and its immediate practical application as inseparable \cite{diniz2024skill,dix2000lifelong}.

Formal Academic Education (T3, $n=35$) emerged with notable prominence despite the skill gap debate. While the tension documented by \citet{diniz2024skill} operates between industry and academia, the individual professionals in this sample perceive formal education as complementary, a foundational layer that enables them to absorb successive waves of change throughout their careers.

Social Learning and Networking (T4, $n=30$) highlights communities of practice, mentoring, and peer exchange, dimensions that in software engineering map onto meetups, open-source communities, and tech talks. Leadership Development and Soft Skills (T5, $n=17$) reflects recognition that behavioral competencies complement technical expertise. The emergent observation (EO1, $n=5$) offers a counterpoint: a small subset favors self-directed paths, signaling a tension between organizational and individual learning agendas.

For organizations, training programs that isolate technical content from practical application, conceptual foundations, and social interaction are likely to be perceived as less useful. Effective strategies should treat training as part of a broader learning ecosystem. Four propositions (PR1--PR4) were derived to guide future quantitative investigations.

For software engineering education, these results suggest that curricula and training programs should not be designed as isolated events but as articulated ecosystems that combine formal foundations, practical application, continuous technical updating, and social learning.

Future work should test these propositions, explore the identified dimensions longitudinally, examine whether the patterns hold across different contexts, and investigate the three-actor dynamic between industry, academia, and the individual professional in the skill gap debate.

\section*{Artifact Availability}\label{sec:artefact}
The replication package for this study includes (i) the anonymized CSV with the open-ended responses and the intermediate artifacts of the six-stage thematic analysis (initial codes, candidate thematic axes, and final theme assignments) and (ii) the complete questionnaire instrument in its original Portuguese version. The package is available at \href{https://doi.org/10.5281/zenodo.19896302}{Zenodo (10.5281/zenodo.19896302)}.

\section*{Acknowledgements}\label{sec:acknowledgements}
We would like to thank all of our participants who dedicated their time to answering the questionnaire. This manuscript was originally drafted in Portuguese. The authors used Generative AI tools (Anthropic Claude) to assist with the translation of the manuscript to English and to improve textual cohesion, clarity, grammar, and engagement of the existing text, following the SBES 2026 policy on the use of AI tools.

\bibliographystyle{ACM-Reference-Format}
\bibliography{sample-base}

%% If your work has appendices, this is the place to put them

\end{document}